\documentclass{aa}
\usepackage{graphicx}
\input{psfig.sty}
\begin{document}
   \title{Structure of visible and dark matter components in spiral galaxies at
   redshifts $z = 0.5 - 0.9$}
   \author{A. Tamm\inst{1} \and
           P. Tenjes\inst{1, 2} }
   \offprints{P. Tenjes}
   \institute{Institute of Theoretical Physics, Tartu University,
              T\"ahe 4, Tartu, 51010 Estonia\\
              \email{atamm@ut.ee; ptenjes@ut.ee}
         \and
             Tartu Observatory, T\~oravere, Tartumaa,
             61602 Estonia}
   \date{Received 23.08.2004; accepted 29.11.2004}

   \abstract{
We have constructed self-consistent light and mass distribution
models for four disk galaxies at redshifts $z=0.48,$ 0.58, 0.81
and 0.88, using the HST archive WFPC2 observations (HDF-S, MDS,
Groth Strip survey) and rotation curves measured by Vogt et al.
(\cite{vogt1}) and Rigopoulou et al. (\cite{rigo}). The models
consist of three components: a bulge, a disk and a dark matter
halo. Similarly to the sample studied in Paper I (Tamm \& Tenjes,
\cite{ta:te2}), light distribution of the galaxies in the outer
parts is clearly steeper than a simple exponential disk. After
applying k-corrections, calculated mass-to-light ratios for
galactic disks within the maximum disk assumption are $M/L_B =$
0.9, 7.4, 4.3 and 1.4, respectively. Together with the galaxies
from Paper I, the mean $\langle M/L_B\rangle =$ 2.5 at $\langle
z\rangle \simeq 0.9$, indicating no significant evolution of
$M/L_B$ with redshift. Central densities of dark matter halos for
an isothermal model are 0.008, 0.035, 0.013, and 0.022 in units
$\rm M_{\sun} /pc^3,$ respectively. Together with the galaxies
from Paper I, the DM central density of the four galaxies at mean
readshift $\langle z\rangle \simeq 0.9$ is $\rho (0) =
(0.012-0.028)\ \rm M_{\sun}/pc^3,$ also showing no significant
evolution with redshift.

We have also constructed mass distribution models without assuming
flat rotation curves and without a dark matter component. Due to the
limited extent of the observed rotation, the models without a dark
halo nearly fit the observations. In this case, mass-to-light ratios
for the galactic disks are 1.8, 9.7, 12. and 1.9, respectively.
   }
\authorrunning{Tamm \& Tenjes}
\titlerunning{spiral galaxies at intermediate redshifts}
   \maketitle
%

\section{Introduction}

The study of dark matter halo central densities and mass
distribution allows us to constrain possible galaxy formation
models and large scale structure formation scenarios (Navarro \&
Steinmetz \cite{na:st}; Khairul Alam et al. \cite{khai}; Gentile
et al. \cite{gent}). For this kind of analysis, it is necessary to
know both the distribution of visible and dark matter.
Unfortunately, the structure and mass distribution of stellar
populations, and therefore, the discrimination between visible and
dark matter in galaxies (especially in very distant galaxies) is
not known precisely enough. Thus, it is interesting to analyze the
corresponding data at different redshifts.

In most cases, the evolution of the visible structure of galaxies
with redshift has been studied, e.g. the luminosity function (LF),
C-M diagram, fundamental plane (FP), the Tully-Fischer (TF)
diagram, scale-lengths of galactic disks.

The luminosity function in deep fields has been studied by Gabasch
et al. (\cite{gaba}) in the FORS Deep Field up to redshifts $z
\sim 5$, Poli et al. (\cite{poli}) in the HDS-S up to $z \sim
3.5,$ and certain evolutionary changes have been found. At the
faint end of the LF, no significant evolution has been detected,
but at the bright end clear brightening was found.  Even at
intermediate redshifts ($z< 1.5$) significant evolution was
detected at the bright end by Cowie et al. (\cite{cowi}). Toft et
al. (\cite{toft}) studied the LF diagram of an X-ray cluster at
$z=1.2$ and in addition to the brightening they found also
flattening of the faint end slope of the LF.

The C-M diagram of a large number of galaxies in the redshift
range $0.2 < z < 1.1$ was studied by Bell et al. (\cite{bell1}),
and evolutionary trends for both early type and late type galaxies
were found. While the early type galaxies evolve only mildly with
redshift, the number of luminous late type galaxies increases
remarkably with redshift.

The fundamental plane in two clusters ($z= 0.58$ and 0.83) was
studied by Wuyts et al. (\cite{wuyt}). They confirm an earlier
result that the slopes of the FP do not change within the studied
redshift range. However, there is a general evolutionary shift
indicating that galaxies in these clusters formed at $z\approx 3.0$.

To study the evolution of the TF diagram, the maximal rotation
velocities in rich cluster galaxies at redshifts $z=0.3-0.5$ were
measured by Ziegler et al. (\cite{zieg1}). Their analysis did not
show any significant difference between cluster and field galaxies
at the same redshifts, even if compared to the local galaxies. At
higher redshifts ($z=0.2-1.2$, mean $z\simeq 0.7$), the TF diagram
was studied on the basis of 83 disk galaxies by Conselice et al.
(\cite{cons}) and also no statistically significant evolution with
redshift was detected. On the other hand, Milvang-Jenses et al.
(\cite{milv}) found that cluster galaxies are brighter by
$0.5^m-1^m$ in B color when compared to field galaxies at the same
redshift ($z=0.8$). For field galaxies at redshifts $z=0.1-1.0$ the
TF diagram was obtained by B\"ohm et al. (\cite{bohm}) and
brightening of galaxies with redshift depending on galactic mass was
discovered.

If, in addition to the surface photometrical data, the rotation
velocities of galaxies are also available, galactic structure can be
described in greater detail. In several projects, the rotation of
distant galaxies has been measured. In addition to the works
referred in Paper I, Erb et al. (\cite{erb}) measured rotation
curves (RCs) of 6 very distant ($z> 2$) galaxies. Unfortunately, the
extent of these RCs is rather small (less than $1''$), the maximum
rotation values have often not been reached and the models of these
galaxies would be uncertain.

In relation to the study of the TF diagram, Ziegler et al.
(\cite{zieg1}) measured RCs for 13 galaxies in rich clusters at
redshifts $z=0.3-0.5.$ Of these galaxies $4 -5$ objects can be
used for mass modeling (the extent of the measured RCs $> 2''$,
kinematics more or less regular). The largest sample of RCs at
intermediate redshifts was measured by J\"ager et al.
(\cite{jage}), where objects in 7 clusters at redshifts
$z=0.3-0.6$ were analyzed and several of the derived RCs could be
used for mass distribution modeling.

To construct a self-consistent photometrical and dynamical model of
a galaxy, both surface photometry and rotation curve data are
needed. For galaxies at intermediate redshifts, the typical scale is
$4-8$~kpc per arcsec and in order to determine the parameters of
galactic components, one should use high-resolution Hubble Space
Telescope (HST) photometry in addition to the other photometrical
data.

In the present study, we continue our modeling of distant galaxies
(Tamm \& Tenjes, \cite{ta:te2}, Paper I). To derive the surface
brightness distribution of these galaxies, we used images from the
HST archive. RCs were measured by Vogt et al. (\cite{vogt1}) and
Rigopoulou et al. (\cite{rigo}). General properties of these
galaxies are given in Table~\ref{tab1}. Galactic names are from the
NED database. In the present work we take $H_0 =$ $\rm 65~km\
s^{-1}Mpc^{-1}$ and $q_0 = 0.5.$

\begin{table*}
   \caption[]{General galactic parameters.}
   \label{tab1}
\begin{center}
\begin{tabular}{llllllllll}
\hline
Name     &  RA          & DEC         &Hubble&$z^a$& Scale &Inclin.$^a$ & $m_I^b$ &$A_B^c$ & $M_B^c$\\
         &  (2000)      & (2000)      & type &     & $\rm (kpc/'')$&(deg)& (mag) & (mag) & (mag)   \\
\hline
GSS 104-4024 &$\rm 14^h17^m26.9^s$&$\rm 52^o24'49.5''$& Sbc  & 0.812 & 6.52 & 82 & 21.88  & 0.034 & $-20.5$\\
GSS 064-4442 &$\rm 14^h17^m54.0^s$&$\rm 52^o29'14.3''$& Sbc  & 0.877 & 6.62 & 55 & 22.05  & 0.033 & $-20.4$\\
MDS uem0-043 &$\rm 3^h4^m59.1^s$&$\rm -0^o11'47.0''$& Sc & 0.476 & 5.51 & 46     & 19.21  & 0.362 & $-21.4$\\
HDFS J223247.66 &$\rm 22^h32^m47.7^s$&$\rm -60^o33'35.9''$& Sb(c)& 0.581& 5.95 & 49 & 19.14   & 0.119 & $-21.2$\\
-603335.9       &                    &                    & & & &
&   & & \\
 \hline
\end{tabular}
\end{center}
\begin{list}{}{}
\item[$^a$] Galactic redshifts are from Vogt et al. (\cite{vogt1})
and (for HDFS J223247.66-603335.9) Rigopoulou et al.
(\cite{rigo}), inclinations are from Simard et al. (\cite{sima})
and Rigopoulou et al. (\cite{rigo}).\\
\item[$^b$] Integrated apparent $I$ magnitudes calculated from our
photometry.\\
\item[$^c$] Absolute $B$-magnitudes in the galactic rest-frame are
calculated from our models (Sect.~4), considering the absorption
in the Galaxy $A_B$ according to Schlegel et al. (\cite{schl}).\\
\end{list}
\end{table*}


\section{Observations and photometry}

We used the HST WFPC2 images of the galaxies for photometry. The
images of GSS 104-4024, GSS 064-4442 and MDS uem0-043 were
retrieved from the HST archive, HDFS J223247.66-603335.9 is a
galaxy on the Hubble Deep Field South, available via the HST web
page. General data about the observational programs and the images
are presented in Table~\ref{tab2}.

\begin{table*}
   \caption[]{HST observations used for photometry.}
   \label{tab2}
\begin{center}
\begin{tabular}{lllll} \hline
Name$^a$ & HST survey & total exposure & background & background\\
         & program    & time in $I$ (s)       & level in $I$ & level in $V$ \\
\hline
GSS 104-4024    & Groth Strip Survey  & 4400 & 5.9   & 6.9  \\
GSS 064-4442    & Groth Strip Survey  & 4400  & 6.15   & 6.9  \\
MDS uem0-043    & Medium Deep Survey  & 6700  & 13.5   & 23.3  \\
HDFS J223247.66-603335.9 & HDF-South  & 113\,900 & 0.0  & 0.0   \\
\hline
\end{tabular}
\end{center}
\end{table*}

\begin{figure*}
\begin{center}
\setlength{\unitlength}{1cm}
\begin{picture}(18,5)
\put(0.0,0.0){\includegraphics{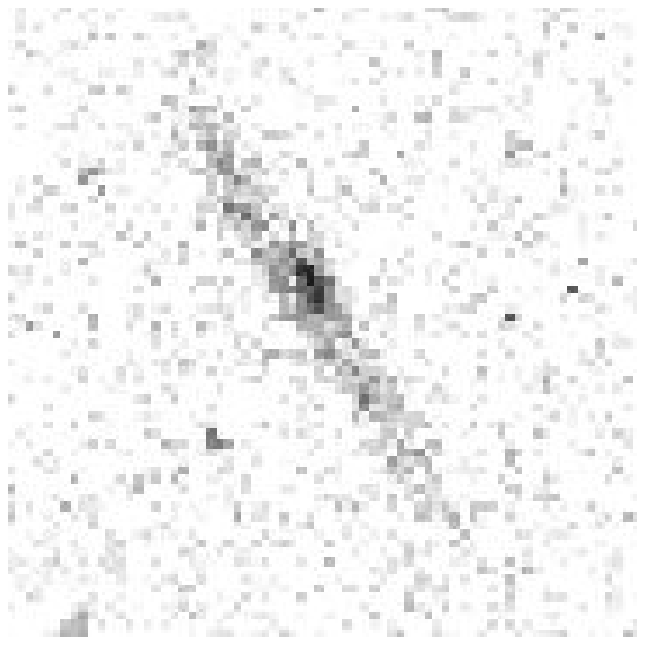}}
\put(9.0,0.0){\includegraphics{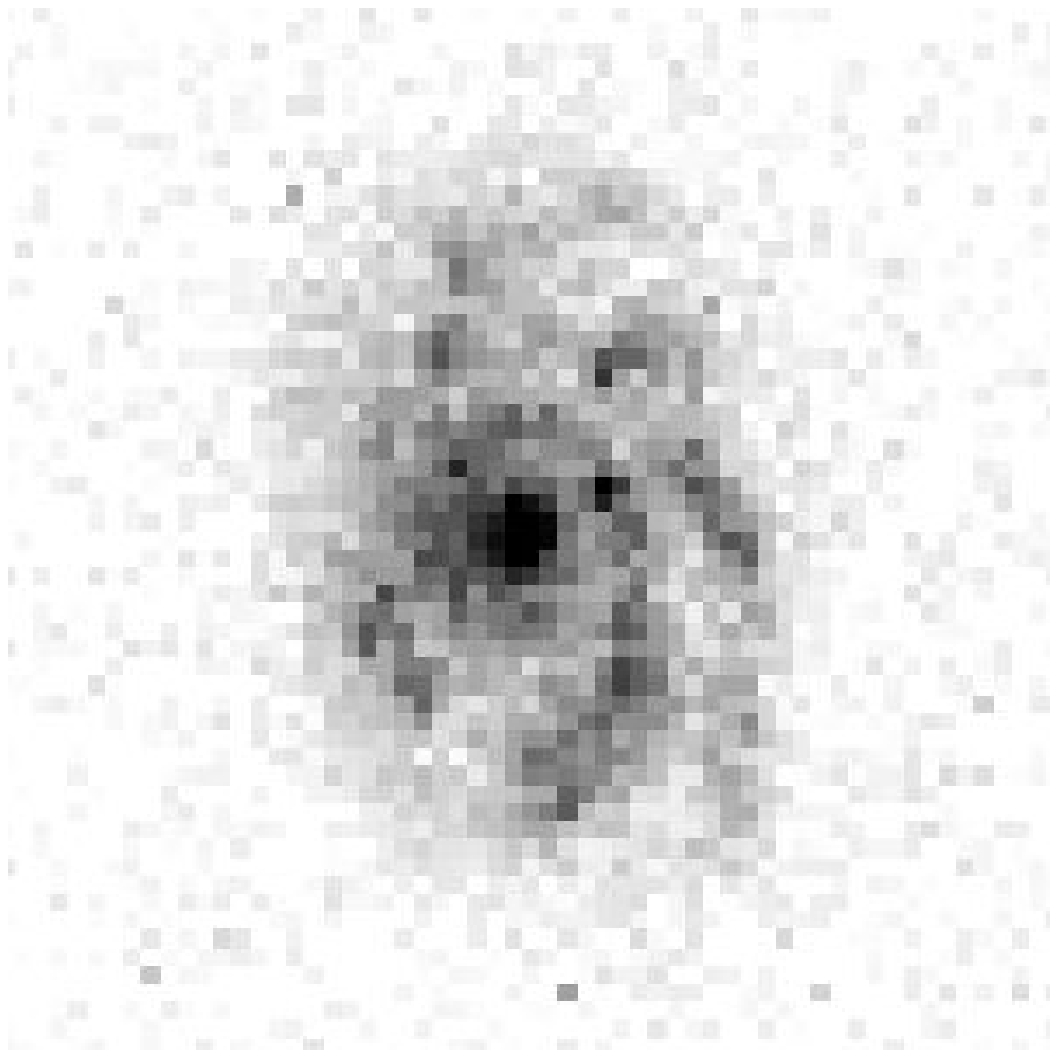}}
\put(4.5,0.0){\includegraphics{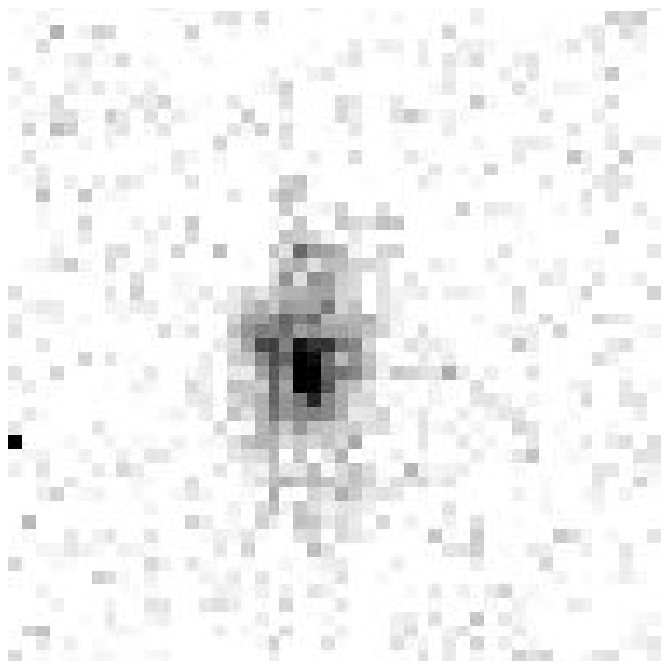}}
\put(13.5,0.0){\includegraphics{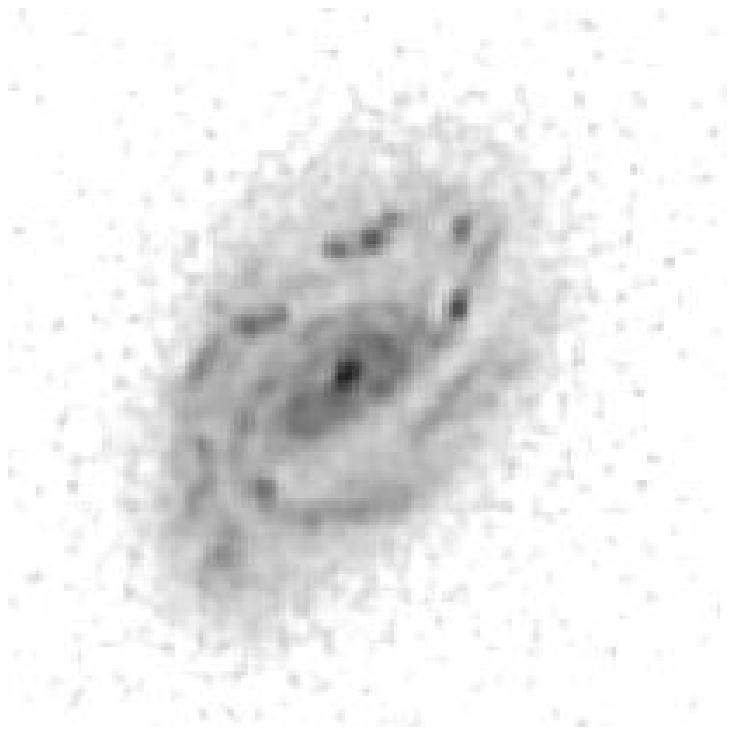}}
\end{picture}
\end{center}
\caption{Images of the galaxies (from left to right) GSS 104-4024,
GSS 064-4442, MDS uem0-043 and HDFS J223247.66-603335.9 through
F814W filter after background subtraction.} \label{fig1}
\end{figure*}

We used pipeline-reduced, on-the-flight calibrated images. The
Hubble Deep Field South is available with all necessary reduction
done, for the other three galaxies we conducted exposure
combining, cosmic-ray removal and background removal with IRAF and
STSDAS software packages, as described in detail in Tamm \& Tenjes
(\cite{ta:te1}, \cite{ta:te2}). The resulting images of the
galaxies can be seen in Fig~\ref{fig1}.

We used the Lucy-Richardson method to deconvolve point spreading. We
calculated point-spread functions of the WFPC2 camera using TinyTim
6.0. The \emph{V}-band image of GSS 104-4024 could not be
deconvolved because of too low signal-to-noise and
signal-to-background ratios. Isophotes were calculated using the
\emph{ellipse} task of STSDAS. To construct luminosity profiles from
the isophotes, we used deconvolved images for the central parts of
galaxies and original images for the outer radii.

\begin{figure*}
\resizebox{\hsize}{!}{\includegraphics*[angle=-90]{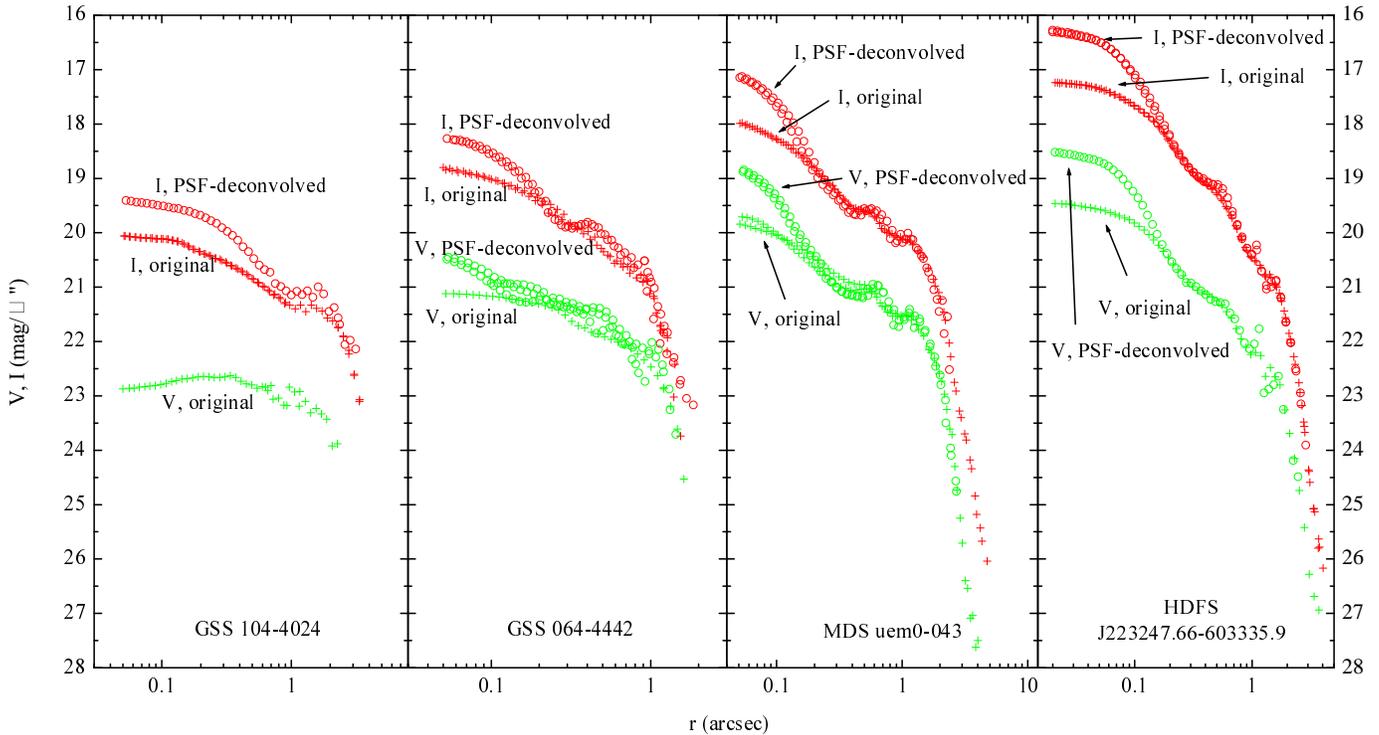}}
\caption{Surface brightness distribution of the galaxies (from
left to right) GSS 104-4024, GSS 064-4442, MDS uem0-043 and HDFS
J223247.66-603335.9 in I and V colors in logarithmic distance
scale. Both original and deconvolved profiles are given.}
\label{fig2}
\end{figure*}

We calibrated WFPC2 luminosity counts to Johnson standard magnitudes
using  formulae and tables given by Holtzman et al (\cite{holt2}),
see also Tamm \& Tenjes (\cite{ta:te1}, \cite{ta:te2}) for details.
Luminosity was corrected for the cosmological dimming by a factor of
$(1+z)^4.$ The resulting luminosity profiles of both original and
deconvolved images in both \emph{V} and \emph{I} passbands are
presented in Fig.~\ref{fig2}. To present luminosity profiles we
prefer a logarithmic scale for distances, enabling a better overview
of the deconvolution results. The scatter of the datapoints of
different runs of the \emph{ellipse} task illustrates the
uncertainties caused by noise and the limited resolution of the
images.

\begin{figure*}
\resizebox{\hsize}{!}{\includegraphics*[angle=-90]{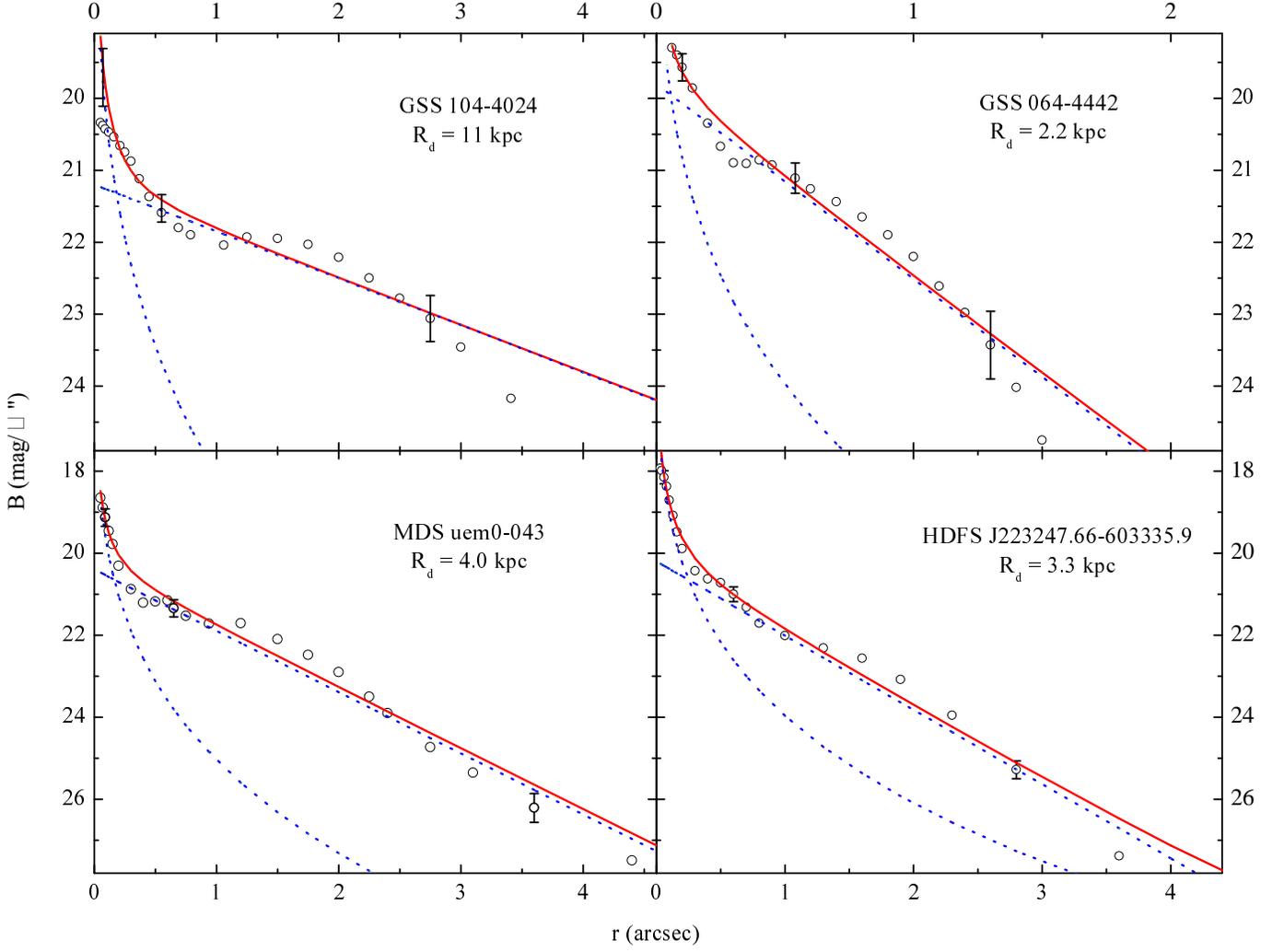}}
\caption{Surface brightness distribution of the galaxies GSS
104-4024, GSS 064-4442, MDS uem0-043 and HDFS J223247.66-603335.9
in rest-frame B color (open circles). Dashed lines -- surface
brightness distribution of best fit $r^{1/4}$ bulge and
exponential disk, solid line -- total surface brightness
distribution.} \label{fig3}
\end{figure*}

\begin{figure*}
\resizebox{\hsize}{!}{\includegraphics*[angle=-90]{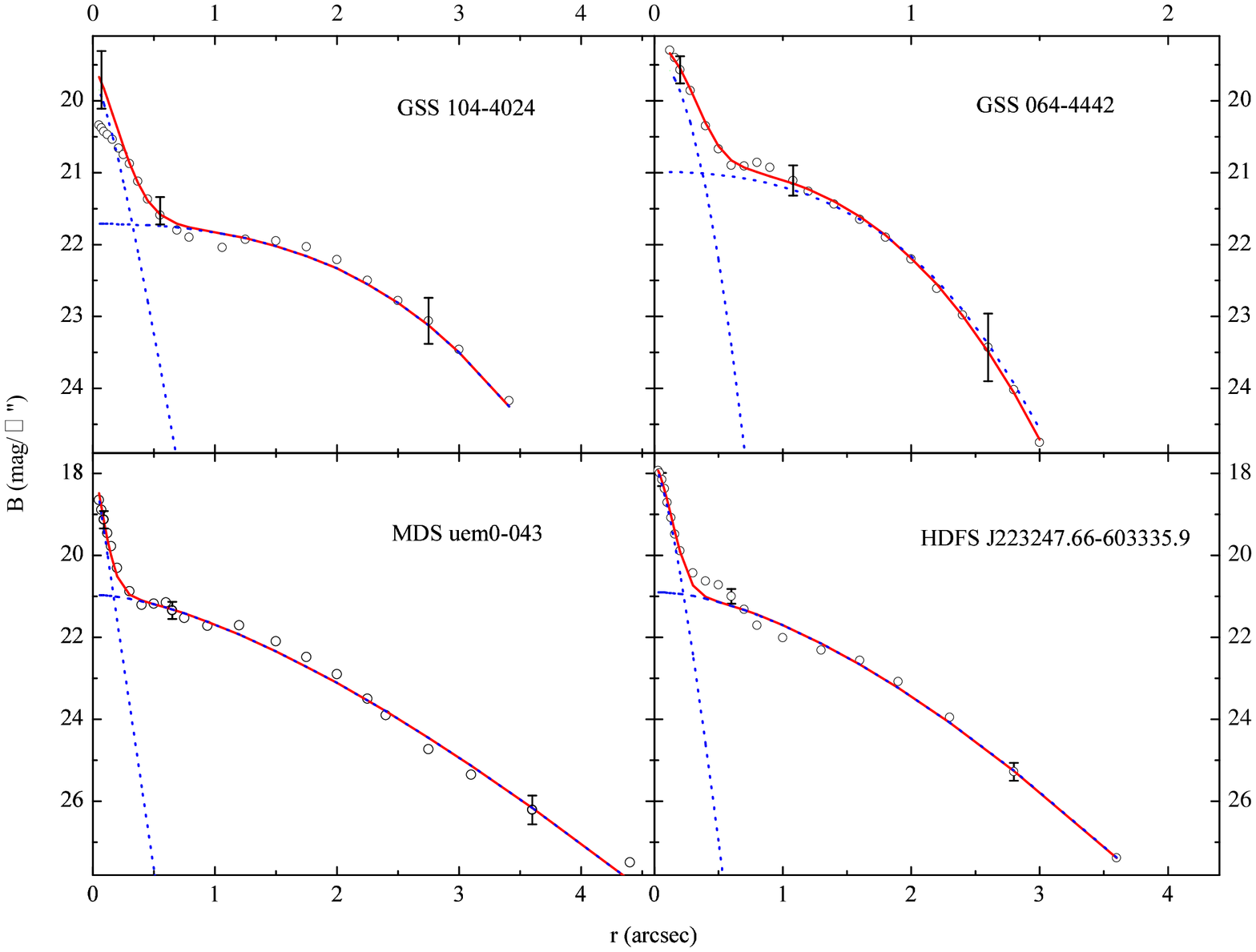}}
\caption{Surface brightness distribution of the galaxies GSS
104-4024, GSS 064-4442, MDS uem0-043 and HDFS J223247.66-603335.9
in rest-frame B color (open circles). Dashed lines -- surface
brightness distribution of best fit model components using spatial
density distribution (\ref{eq2}). Solid line -- total surface
brightness.} \label{fig4}
\end{figure*}

For modeling we derived rest-frame \emph{B}-band luminosities,
applying the k-correction according to van Dokkum \& Franx
(\cite{do:fr1}), Kelson et al. (\cite{kels2}). Synthetic spectra
were taken from the paper by Coleman et al. (\cite{cole}), the
relation between Johnson and $AB$-magnitudes from Frei \& Gunn
(\cite{fr:gu}). We determined the following transformations to be
conducted
$$\begin{array}{ll}
B(z=0.48) & = I + 0.585(V-I) + 0.7\\
B(z=0.58) & = I + 0.36(V-I) + 0.86\\
B(z=0.81) & = I + 0.95\\
B(z=0.88) & = I - 0.106(V-I)+1.18\\
\end{array} $$
The resulting luminosity profiles in the galactic rest-frame
\emph{B} color are shown in Figs~\ref{fig3} and \ref{fig4} by open
circles. These profiles were used for further modeling.

The RCs for the galaxies GSS 104-4024, GSS 064-4414 and MDS
uem0-043, based on Keck spectroscopy, were taken from the paper by
Vogt et al. (\cite{vogt1}). The velocity profile of HDFS
J223247.66-603335.9 was measured by Rigopoulou et al. (\cite{rigo})
using VLT spectra.

We have minimized the difference between the RCs from the two sides
of each galaxy, shifting the coordinates $(r_0, V_0)$ of the RC
panels (see Persic \& Salucci (\cite{pe:sa}) for details). This
makes the RC profiles more regular. The resulting folded RCs are
presented with circles in Figs ~\ref{fig5}$-$~\ref{fig8}. We have
used different types of circles (open and closed) to discriminate
between the observations in opposite directions from the center.
Error bars are also shown. We have tried to select galaxies with
kinematics as regular as possible.

\section{Model description}
As in Paper I, we limit the main stellar components to the bulge
and the disk. To construct a dynamical model, a dark matter
component -- the dark halo -- must be added to visible components.

To construct a model with two stellar components, the surface
luminosity distribution is usually approximated by a $r^{1/4}$-bulge
and an exponential disk. These surface brightness distributions can
be expressed with the help of the general S\'{e}rsic formula
(\ref{eq1})
\begin{equation}
I(a) = I(0) \exp [ -b_m (a/a_c)^{1/m}], \label{eq1}
\end{equation}
(S\'{e}rsic \cite{sersic}) with a parameter $m\ge 2$ for the bulge
and $m=1$ for the disk. In this formula $a$ is distance along the
galactic major axis, $a_c$ is the radius containing half of the
total luminosity, $b_m$ is a normalizing constant.

If, in addition to the photometrical data, kinematic data are also
used, the corresponding dynamical model must be consistent with the
photometry, i.e. the same density distribution law must be used for
RC modeling (and for the velocity dispersion curve, if possible). In
the case of an infinitely thin disk, the circular velocity for an
exponential surface density distribution can be expressed via
modified Bessel functions (Freeman \cite{freeman}). In addition, for
spherical systems an expression for circular velocity with an
integer S\'{e}rsic index $m$ (Mazure \& Capelato \cite{ma:ca}) can
be derived. For a non-integer index and ellipsoidal surface density
distribution a consistent solution for RC calculations is not known.

The density distribution parameters are often determined by the
least squares method. In this case the parameter $m$ is not
necessarily an integer number. In the present paper, the density
distribution parameters are determined by the least squares method
and they can have any value. Moreover, we have found that at least
for our present sample of high redshift galaxies the best fit for
the disk luminosity distribution is achieved with the S\'{e}rsic
index $m<1$.

For the reasons given above we decided in addition to models with a
$r^{1/4}$ bulge and exponential disk to construct models for the
galaxies starting from a spatial density distribution law, which
allows an easier least-square fitting simultaneously for light
distribution and RC.

In such models, the visible part of a galaxy is given as a
superposition of a bulge and a disk. The spatial density
distribution of each visible component is approximated by an
inhomogeneous ellipsoid of rotational symmetry with the constant
axial ratio $q$ and the density distribution law
\begin{equation}
\rho (a)=\rho (0)\exp [ -( a/(ka_0))^{1/N} ] , \label{eq2}
\end{equation}
where $\rho (0)=hM/(4\pi q a_0^3)$ is the central density and $M$ is
the component mass; $a= \sqrt{R^2+z^2/q^2}$, where $R$ and $z$ are
two cylindrical coordinates. $a_0$ is the harmonic mean radius which
characterizes rather well the real extent of a component,
independently of the parameter $N$. Coefficients $h$ and $k$ are
normalizing parameters, depending on $N$, which allows the density
behavior to vary with $a$. The definition of the normalizing
parameters $h$ and $k$ and their calculation is described in Tenjes
et al. (\cite{tenj1}). Equation (\ref{eq1}) allows a sufficiently
precise numerical integration and has a minimum number of free
parameters.

The dark matter (DM) distribution is represented by a spherical
isothermal law
\begin{equation}
\rho (a) = \cases{\rho (0) ([1+({a\over a_c})^2]^{-1} -
                      [1+({a^0\over a_c})^2]^{-1}) & $a \leq a^0$
                      \cr
                    0         &      $a>a^0$.    \cr }
\label{eq3}
\end{equation}
Here $a^0$ is the outer cutoff radius of the isothermal sphere,
$a_c = ka_0.$

The density distributions for the bulge and the disk were
projected along the line of sight, divided by their
mass-to-luminosity ratios $f$ and their sum gives us the surface
brightness distribution of the model
\begin{equation}
L(A) = 2 \sum_{i=1}^2 {q_i\over Q_i f_i} \int_A^{\infty}
          {\rho_i (a) a~da\over (a^2 - A^2)^{1/2}} ,
\label{eq4}
\end{equation}
where $A$ is the major semiaxis of the equidensity ellipse of the
projected light distribution and $Q_i$ are their apparent axial
ratios $Q^2=\cos^2\gamma+q^2\sin^2\gamma$. The angle between the
plane of a galaxy and the plane of the sky is denoted by $\gamma$.
The summation index $i$ designates two visible components, the
bulge and the disk.

The masses of the components were determined from the rotation law
\begin{equation}
v_i^2(R) = 4\pi q_i G \int_{0}^{R} {\rho_i (a) a^2 da\over
              (R^2-e_i^2a^2)^{1/2}},
\label{eq5}
\end{equation}
\begin{equation}
   V^2(R) = \sum_{i=1}^3 v_i^2(R),
\label{eq6}
\end{equation}
where $G$ is the gravitational constant, $e=\sqrt{1-q^2}$ is
eccentricity, and $R$ is the distance in the equatorial plane of
the galaxy. Now the summation is over all three components.

We have kept the axial ratio of the components fixed, taking $q =
0.7$ for the bulge and $q = 0.1$ for the disk, according to analogy
with the near-by galaxies. The model parameters $a_0$, $L_B$, $M$
and $N$ for the bulge and the disk were determined by a subsequent
least-squares approximation process. First, we made a crude
estimation of the population parameters. The purpose of this step is
to avoid obviously non-physical parameters -- relations (\ref{eq4})
and (\ref{eq5}) are nonlinear and fitting of the model to the
observations is not a straightforward procedure. Next, a
mathematically correct solution was found for each galaxy. Details
of the least squares approximation and the general modeling
procedure were described by Einasto \& Haud (\cite{ei:ha}), Tenjes
et al. (\cite{tenj1}, \cite{tenj2}).

The RCs of the modeled galaxies have a rather small extent (see
Figs. \ref{fig5}--\ref{fig8}). We do not have any additional
information about the dynamics of these galaxies at large
galactocentric radii. For this reason, the outer cutoff radius of
the DM component remains undetermined at present. As in Paper I,
we fixed $a^0 = 5 a_0$ on the basis of the structure of nearby
galaxies. This value influences the behavior of the RC only in the
extreme outer regions and does not influence our results. In
addition, by analogy with nearby galaxies we assumed that the RCs
of these galaxies remain nearly flat at least up to 30~kpc $\simeq
1.5-2 R_{opt}$ (Persic et al. \cite{pers}, Sofue \& Rubin
\cite{so:ru}).

As we have no information about velocity dispersions in the
central regions of our galaxies, we can calculate only the
circular velocities (Eqs.~\ref{eq5}--\ref{eq6}). The difference
between the circular and rotational velocities is known in
galactic dynamics as the asymmetric drift and is mainly a function
of velocity dispersions. Typical emission-line dispersions in disk
galaxies at intermediate redshifts are 30--100 km/s or even more
(Im et al. \cite{im1}, Erb et al. \cite{erb}). Thus, within the
central $0.5\arcsec - 0.7\arcsec$, modeled velocities must be
higher than the observed rotational velocities (see also Sect.~4).

\section{Results}

We have fitted the two-component models of the galaxies to the
\emph{B} color profiles. Profiles fitted by a $r^{1/4}$-bulge and an
exponential disk can be seen in Fig.~\ref{fig3}. Profiles fitted on
the basis of the spatial density law (\ref{eq2}) are presented in
Fig.~\ref{fig4}. In both cases dotted lines correspond to the
individual components -- the bulge and the disk. The solid line is
the sum of the two components. It is seen that a non-exponential
disk gives a better agreement with observations. The mean deviations
of the models from the observations in Fig.~\ref{fig4} are only
$\langle \mu^{\rm obs} -\mu^{\rm model} \rangle =$ $0.07^m$,
$0.04^m$, $0.14^m$ and $0.13^m$, respectively.

Fitting of the modeled RCs to the observations is much more complex.
First, the resolution of the present-day spectroscopic observations
is insufficient to construct high-precision velocity profiles for
galaxies at such high redshifts. In addition to the uncertainties
and the low extent of the observations, the velocities in the
central regions remain severely underestimated due to the lack of
resolution and therefore the bulge mass remains practically unknown.

It is possible to fit only the share of the disk or the dark matter
component to the observed rotation and still achieve satisfactory
concordance. In order to keep the results more realistic we have,
however, constructed three-component models for all four galaxies
with a bulge, a disk and a dark matter component. For the reasons
given above, the mass of the bulge component should not be taken too
literally; we have practically ignored the observed velocities at
low radii. As in most cases no other hint of the mass of the bulge
was given, we took the mass-to-light ratio of the bulge equal to
that of the disk. Assuming the bulge of a galaxy to be populated by
stars from an older generation than the disk, we should thereby get
the approximate lower limit of the bulge mass.

We have scaled the original observed RCs to match $V_{\rm max}$
given by the authors, thereby compensating for the effects of
misalignment of the slit, ``beam smearing'' and seeing at the outer
radii, which we have used to fit our modeled RC.

The parameters of the DM component cannot be determined uniquely by
our method. We could, for example, increase its mass, if we
increased \emph{r} simultaneously. A parameter that can be
determined reasonably well, however, is the central density of the
dark matter $\rho (0)$ (see Sect.~5).

\subsection{Notes about individual galaxies}
\subsubsection{GSS 104-4024}

This Sbc galaxy at $z=0.8116$ is seen nearly edge-on (inclination
$82\degr$); a possible dust lane along its semi-major axes can be
detected on the HST image (Fig.~\ref{fig1}), obscuring the central
part of the bulge. The usual isophote fitting would underestimate
the luminosity of the central regions ($r < 0.3\arcsec$); the true
luminosity remains unknown. A possible correction suggested by our
modeling software is shown in Fig. ~\ref{fig4} (the correction
contributes only 0.1$^m$ to the total luminosity).

We calculated the total apparent $V$-band luminosity of the galaxy
to be $m_I = 21.88\pm 0.06,$ which matches well with the estimate
22.05 by Simard et al. (\cite{sima}). The mean apparent color index
$(V-I) = 2.1\pm 0.4$. Taking into account the absorption in the
Galaxy, we found the absolute $B$-band luminosity to be $M_B =
-20.5$. However, considering also the internal absorption, the
galactic luminosity would be higher and the resulting $M/L$ ratio of
visible matter would be smaller. For the exponential bulge we find
$r_{d}$ = 11 kpc, which coincides with the value 11.54 given by
Simard et al. (\cite{sima}).

\subsubsection{GSS 064-4442}

This is an Sbc galaxy at $z=0.877$, the most distant one discussed
in this paper. We calculated the total apparent Johnson $m_I =
22.05\pm 0.14,$ which matches well with the estimated value 21.97 by
Simard et al. (\cite{sima}). The mean apparent color index $(V-I) =
1.5\pm 0.2$, absolute $B$-band magnitude $M_B= -20.4$. We obtained
$r_{d}$ = 2.2 kpc for the exponential disk, which somewhat differs
from the estimate 3.76 by Simard et al. (\cite{sima}). Taking into
account the faintness and low resolution of the image of the galaxy
this deviation is not surprising.

\subsubsection{MDS uem0-043}

This well-exposed Sc galaxy at $z=0.476$ has an obvious spiral
structure. We found the total Johnson \emph{I}-band luminosity to be
$19.21\pm0.04,$ which agrees well with the result of Vogt et al.
(\cite{vogt1}) $m_I = 19.31$. The mean apparent color index $(V-I) =
1.5\pm0.2$. With correction for the absorption in the Galaxy
applied, MDS uem0-043 turns to be a bright galaxy: $M_B= -21.4$. We
estimated $r_{d}$ = 4.0 kpc for the exponential disk, which is
reasonably close to $r_{d}$ = 5.24 by Vogt et al. (\cite{vogt1}).
However, in Vogt et al. (\cite{vogt1}) deconvolution of the surface
brightness profile has not been done, thus a slight difference in
disk scale is normal.

\subsubsection{HDFS J223247.66-603335.9}

This bright and massive Sb galaxy is very well exposed on the
HDF-South image and was studied in a paper by Rigopoulou et al.
(\cite{rigo}). The excellent quality of the HDF images makes it
possible to analyze the photometry of the galaxy with high
precision. We calculated the total apparent Johnson \emph{I}
luminosity to be $19.14\pm0.02$. Transforming this result to the
\emph{AB} system (adding 0.35 mag, Frei \& Gunn, \cite{fr:gu}), we
get a perfect match with the estimate of 19.5 by da Costa et al.
(\cite{cost}) on the basis of ground-based observations with the
VLT. The mean apparent color index $(V-I)= 2.0\pm0.2$. The absolute
B-band luminosity is $M_B= -21.2$. We obtained $r_{d}$ = 3.3 kpc for
the exponential disk, which unfortunately cannot be compared to the
value by Rigopoulou et al. (\cite{rigo}) due to different surface
brightness scaling.

The velocity profiles are presented in Fig.~\ref{fig8}. Calculated
from our model, the total mass within 28~kpc is $13\cdot 10^{11}\rm
M_{\sun}$, which agrees well with the mass $10\cdot 10^{11}\rm
M_{\sun}$ within 20~kpc estimated by Rigopoulou et al.
(\cite{rigo}).

\subsection{Calculated model parameters}

In Table \ref{tab3}, calculated from the models, the integrated
parameters of a S\'{e}rsic surface density distribution are given:
the scale lenghts of the bulge $r_b$ and the disk $r_d$ and the
corresponding central luminocities in rest-frame $B$-band
$(\mu^{0}_B)_{\rm bulge},$ $(\mu^{0}_B)_{\rm disk}$. The
corresponding profiles can be seen in Fig.~\ref{fig3}.

\begin{table}
   \caption[]{Derived parameters for a S\'{e}rsic surface density distribution.}
   \label{tab3}
\begin{tabular}{lllll}
\hline
Name$^a$        & $r_b$ & $r_d$      & $(\mu^{0}_B)_{\rm bulge}$ & $(\mu^{0}_B)_{\rm disk}$ \\
                & (kpc) &(kpc)& (mag)&  (mag) \\
\hline
GSS 104-4024    & 2.0 & 11.0                 & 14.0                & 21.2   \\
GSS 064-4442    & 0.3 & 2.2                 & 14.5                & 19.8   \\
MDS uem0-043    & 0.23 & 4.0                 & 13.0                & 20.4   \\
HDFS J223247.66 & 0.3 & 3.3                 & 12.7                & 20.2    \\
    -603335.9   &       &       &       &         \\
\hline
\end{tabular}
\end{table}

\begin{figure}
\resizebox{\hsize}{!}{\includegraphics*{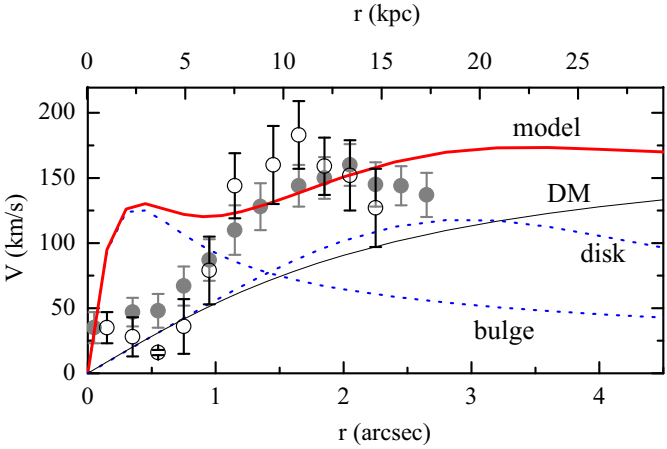}}
\caption{Rotation curve of the galaxy GSS 104-4024. Open and gray
circles -- observed velocities by Vogt et al. (1996) at the
opposite sides of the galaxy, thick solid line -- rotation curve
from the mass model, dashed lines -- rotation velocities due to
visible components, thin solid line -- rotation velocities due to
the dark matter component.} \label{fig5}
\end{figure}

\begin{figure}
\resizebox{\hsize}{!}{\includegraphics*{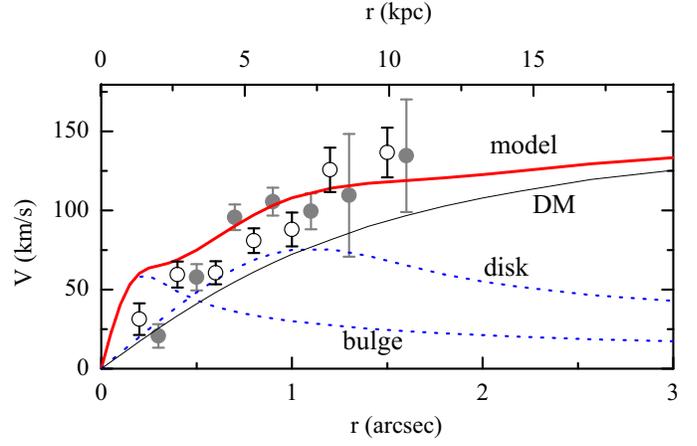}}
\caption{Rotation curve of the galaxy GSS 064-4442. Designations
are the same as in Fig.~\ref{fig5}.} \label{fig6}
\end{figure}

\begin{figure}
\resizebox{\hsize}{!}{\includegraphics*{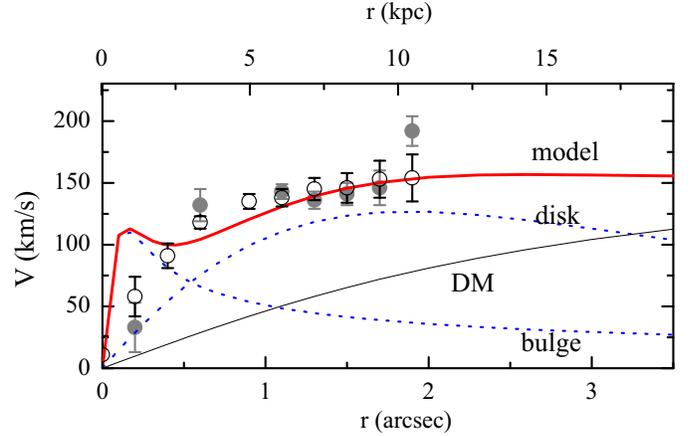}}
\caption{Rotation curve of the galaxy MDS uem0-043. Designations
are the same as in Fig.~\ref{fig5}.} \label{fig7}
\end{figure}

\begin{figure}
\resizebox{\hsize}{!}{\includegraphics*{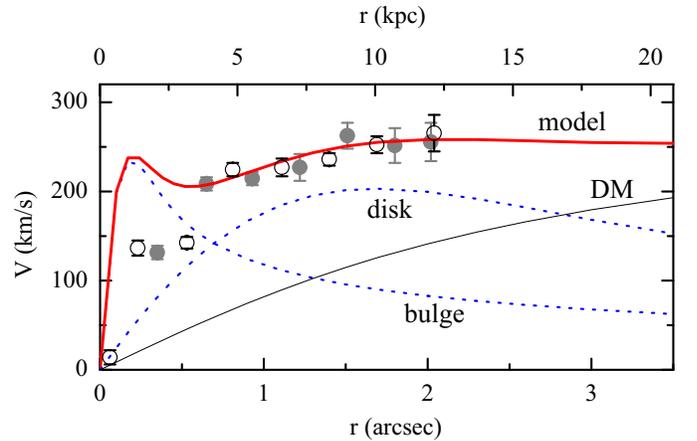}}
\caption{Rotation curve of the galaxy HDFS J223247.66-603335.9.
Designations are the same as in Fig.~\ref{fig5}.} \label{fig8}
\end{figure}

The parameters of the final models on the basis of the spatial
density distribution (\ref{eq2}) (the harmonic mean radius $a_0$,
the total mass of the population $M$, the structural parameter
$N$, the dimensionless normalizing constants $h$ and $k$,
B-luminosities and the corresponding mass-to-light ratios) are
given in Table~\ref{tab4}. The masses of the bulge and DM are
given in parentheses to indicate that on the basis of the present
observational data it is not possible to determine these values
uniquely. The final model profiles are denoted by thick solid
lines in Figs.~\ref{fig4}--\ref{fig8}. Models of individual
visible components are presented by dashed lines, the model of the
dark matter component by a thin solid line.

\begin{table*}
   \caption[]{Model parameters for spatial density distribution.}
   \label{tab4}
\begin{center}
\begin{tabular}{llllllll}
\hline
Name &$a_0$&  $M$                    &$M/L_B$& $L_B$
& $N$ & $h$ & $k$ \\
     &(kpc)&$(\mathrm 10^{10}M_{\sun})$& $(M_{\sun}/L_{\sun})$ &
$(\mathrm 10^{10} L_{\sun})$ &     &     & \\
    \hline
GSS 104-4024&   &                  &       &    &     &      &       \\
bulge & 1.5 & (2.2)               & 4.3 & 0.50    &0.91&3.33&0.594  \\
disk  & 13.2& 8.7                & 4.3  & 2.03  &0.32&1.18&1.37  \\
dark matter& 80. & (550.)               & &      &    &14.8&0.151\\
GSS 064-4442&  &                     &       &    &     &      &       \\
bulge & 0.9 & (0.3)               & 0.90  & 0.31  &0.51&1.59&1.12 \\
disk  & 5.0 & 1.8                 & 0.90  & 2.03  &0.34&1.23&1.34 \\
dark matter& 50. & (230)                &    &   &    &14.8&0.151 \\
MDS uem0-043&  &                     &       &    &     &      &       \\
bulge & 0.5 & (0.6)               & 1.41  & 0.41  &1.11&4.96&0.405 \\
disk  & 7.2 & 7.1                 & 1.41  & 5.02 &0.73&2.38&0.805  \\
dark matter& 80. & (340.)               & &       &    &14.8&0.151 \\
HDFS J223247.66& &          &       &    &     &      &       \\
-603335.9 & & & & & & & \\
bulge & 0.7 & (5.6)               & 7.4  & 0.75  &0.80&2.70&0.719  \\
disk  & 7.0 & 28.3                & 7.4   & 3.83  &0.67&2.12&0.888 \\
dark matter& 80. & (1500.)             & &       &    &14.8&0.151 \\
\hline
\end{tabular}
\end{center}
\end{table*}

In Table~\ref{tab5} the total mass of the visible matter $M_{\rm
vis}$, the total intrinsic luminosity $L_B$, the mass-to-light
ratio of the visible matter $M/L_B$ and the disk-to-bulge ratio
$D/B$ are given.

\begin{table}
   \caption[]{Derived integrated parameters from models.}
   \label{tab5}
\begin{tabular}{lllll}
\hline
Name            & $M_{\rm vis}$       & $L_B$               & $M/L_B$ & $D/B$ \\
                &$\rm 10^{10}M_{\sun}$&$\rm 10^{10}L_{\sun}$&        & \\
\hline
GSS 104-4024    & 11.                 & 2.53                & 4.3   & 4.0 \\
GSS 064-4442    & 2.1                 & 2.34                & 0.9   & 6.5 \\
MDS uem0-043    & 7.6                 & 5.43                & 1.4   & 12.2 \\
HDFS J223247.66 & 34.                 & 4.58                & 7.4   & 5.1 \\
    -603335.9   &        &       &         \\
\hline
\end{tabular}
\end{table}

\section{Discussion}

We have used the isothermal approximation for the dark matter
density distribution in our models. However, several other dark
matter distributions have been suggested and used by different
authors. The main debate has been between "cuspy" distributions
deduced from N-body simulations, with the density increasing
infinitely towards the center of the halo, and "cored"
distributions usually derived from the observed RCs of late type
and low surface brightness galaxies with a constant density
distribution in the central part of the halo.

According to their numerical simulations, Navarro, Frenk \& White
(\cite{nava1}, NFW) proposed the following formula for the density
distribution
\begin{equation}
\rho(r) = {\rho_{s}\over(r/r_{s})[1+(r/r_{s})]^{2}}, \label{eq7}
\end{equation}
where $\rho_{s}$ and $r_{s}$ are the characteristic density and
radius, respectively. Moore et al. (\cite{moor}) proposed a rather
similar, but still "cuspier" distribution:
\begin{equation}
\rho(r) = {\rho_{sM}\over(r/r_{s})^{1.5}[1+(r/r_{s})^{1.5}]}
\label{eq8}
\end{equation}

Burkert (\cite{burkert}) studied low surface-brightness galaxies
and approximated the dark component with a distribution similar to
isothermal
\begin{equation}
\rho(r) = {\rho_0 r_0^3\over(r+r_0)(r^2+r_0^2)} \label{eq9}
\end{equation}

\begin{figure}
\resizebox{\hsize}{!}{\includegraphics*{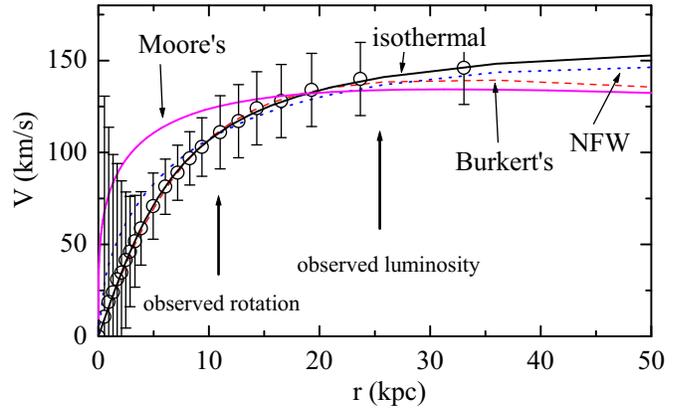}}
\caption{Comparison of the dark matter contribution in kinematics of
the galaxy MDS uem0-043 (open circles with estimated error bars)
with the different dark matter density distribution models. Two
vertical arrows indicate outermost observed points of the surface
brightness profile and the rotation curve.} \label{fig9}
\end{figure}

We have constructed a "best estimate" curve for the velocity
distribution of the galaxy MDS uem0-043, using the observed rotation
for the inner radii and keeping the rotation velocities beyond them
constant. Thereafter we subtract the share of the modeled visual
components from the resulting curve, revealing an estimate for the
share of the DM component $V^2_{\rm DM} = V^2_{\rm obs} - V^2_{\rm
vis}$. These velocities are shown as open circles in Fig.~\ref{fig9}
together with the estimated errors and they correspond to the
supposed rotation contribution of DM. Finally, we test how well
different density distributions fit with the derived DM
contribution. The large uncertainties at small radii are caused by
the effects of limited spectral and spatial resolution.

Except for the "cuspiest" one by Moore et al. (\cite{moor}), the
curves of all other distributions fit within the error bars
(Fig.~\ref{fig9}). Burkert's model is nearly the same as our
isothermal law in the central part, while the best fit of NFW
stays slightly above these two at small radii and below them at
higher radii. Due to the limited spatial resolution of the RCs of
high redshift galaxies, it is not possible to prefer the NFW,
Burkert or isothermal distributions, nor to study the most recent
profiles by Navarro et al. (\cite{nava2}) and Fukushige et al.
(\cite{fuku}). We intend do this analysis on the basis of detailed
models of nearby well-studied galaxies.

As the behavior of the density distributions (7) and (8) at the
center is limited by the resolution of N-body calculations and the
behavior of density distributions (3) and (9) is quite similar
(Fig.~\ref{fig9}), below we discuss only the isothermal DM halo.

Due to the limited extent of RCs it is not possible to determine
separately the radius and the mass of the DM distribution.
However, the combination of these two parameters, the central
density of the dark matter $\rho (0) = hM/(4\pi a_0^3),$
characterizing the central rotation velocity gradient caused by
the dark halo, is nearly independent of the choice of the dark
halo outer cutoff radius and can be determined. Calculated from
our models, the central densities of DM components are $\rho (0)
=$ 0.013, 0.022, 0.008 and 0.035 in units $\rm M_{\sun} / pc^3$.
These values are calculated for maximum disk models and are thus
the lower limits. Adding the two galaxies at redshifts 0.9 and
0.99 from Paper I we get the range of the DM central density of
the four galaxies at the mean readshift $\langle z\rangle \simeq
0.9$ is $\rho (0) = (0.012-0.028)\ \rm M_{\sun}/pc^3$. This value
can be compared with the mean central DM density of nine local
galaxies $\rm \rho (0) = (1-4)\cdot 10^{-24}g/cm^3=$
$(0.015-0.050)\ \rm M_{\sun}/pc^3$ derived by Borriello \& Salucci
(\cite{bo:sa}). Although our sample of galaxies is small, there
seems to be no significant evolution of DM central density with
redshift.

We also calculated the amount of DM within a half-light radius. For
galaxies at redshifts $\langle z\rangle \simeq 0.9$ the mean dark
matter contribution was $40\pm 15$ per cent. Thus in our case, the
discrepancy between the observed galaxies and CDM models is not as
high as was indicated in the analysis by Bell et al. (\cite{bell2}).

According to our models, the mass-to-light ratios of visible matter
(within the maximum disk assumption) are $M/L_B =$ 4.3, 0.9, 1.4 and
7.4. Adding the two galaxies at redshifts 0.9 and 0.99 from Paper I
to these galaxies, we find that the mean $M/L_B$ of the four
galaxies at the mean redshift $\langle z\rangle \simeq 0.9$ is
$\langle M/L_B \rangle =$ 2.5. Another way to determine stellar
masses and mass-to-light ratios is from broadband optical-NIR
photometry and chemical evolution models (Drory et al.
(\cite{dror}), Berta et al. (\cite{bert})). For galaxies at
redshifts $z\sim$ 0.9 Drory et al. (\cite{dror}) derived $\langle
M/L_B\rangle =$ 1.3, being somewhat smaller than the value
calculated by us. For the highest mass galaxies at redshift $z\sim$
0.5, they derived $\langle M/L_B\rangle =$ 2.6, also smaller than
our value of 7.4 for the galaxy HDFS J223247.66-603335.9. However,
we think that within the precision of the models used these results
agree reasonably. Stellar masses of starburst galaxies at $z=0.5$
are found to be even as large as $\rm 50\cdot 10^{10}M_{\sun}$
(Drory et al. (\cite{dror}), Berta et al. (\cite{bert})).

The derived $M/L_B = 2.5$ at $z\sim 0.9$ can be compared to the mean
stellar $M/L_B$ calculated by Giraud (\cite{giraud}) for local Sb
galaxies $M/L_B = 2.7$. No significant evolution with redshift is
detected -- a result which was found earlier in the studies of the
TF diagram at intermediate redshifts (Vogt et al. \cite{vogt1},
Ziegler et al. \cite{zieg1}, Conselice et al. (\cite{cons}), B\"ohm
et al. (\cite{bohm}, for higher mass galaxies). An explanation for
these small changes can be found in chemical evolution models.
Evolution calculations by Bicker et al. (\cite{bick}, Fig.~7)
indicate that the k-correction and evolution corrections in the
$B$-band nearly compensate each other in the case of Sb galaxies at
redshifts $z\simeq 0.5-1$.

As mentioned in Paper I, the disk concentration parameter N seems
to have a trend towards lower values at higher redshifts. Even
visual inspection of the luminosity profiles (Figs ~\ref{fig2} --
~\ref{fig4}) reveals that the surface brightness of the outer
parts of the galaxies decreases faster than exponentially. Surface
brightness distribution of disks with the concentration parameter
$N < 1$ is similar to the surface brightness distribution of
truncated disks described by Pohlen et al. (\cite{pohl1},
\cite{pohl2}). According to recent SPH simulations, truncated
disks seem to be a natural phenomena within $\Lambda$CDM models
(Governato et al. \cite{gove}, Fig.~2). However, in our case the
truncation takes place generally slightly earlier than at 3--5
disk scale lengths. Truncation of a sample of redshift $z\sim 1$
disk galaxies has recently been studied by P\'{e}rez
(\cite{pere}).

\begin{figure*}
\resizebox{\hsize}{!}{\includegraphics* {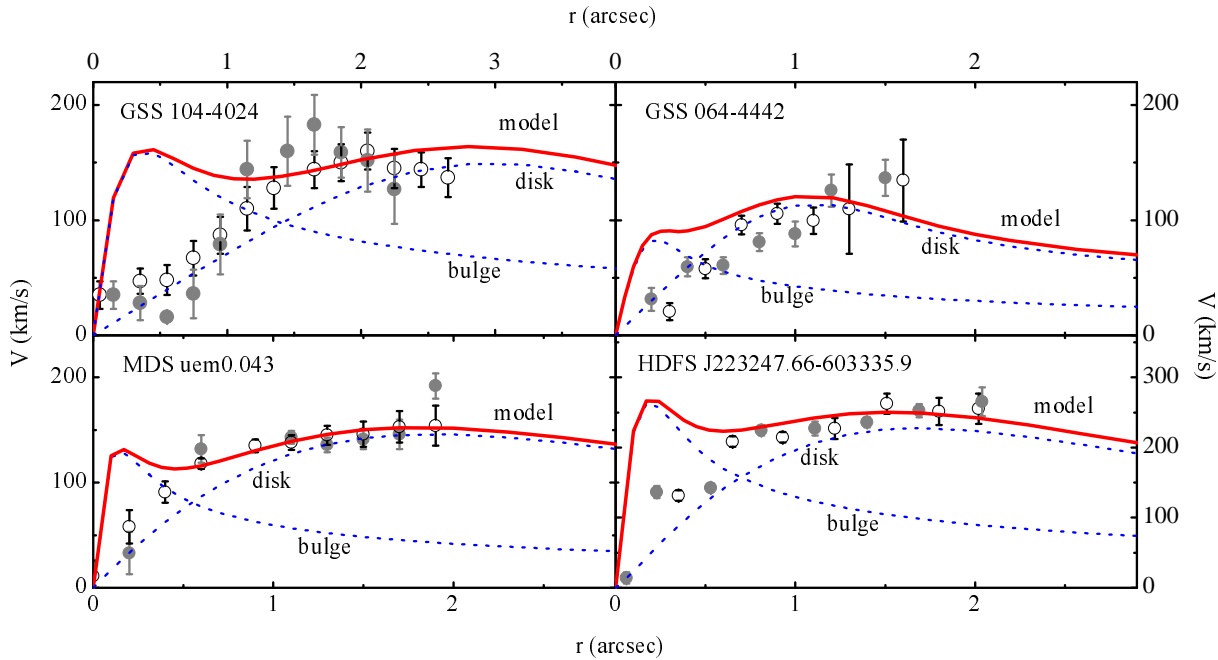}}
\caption{Surface brightness distribution of the galaxies GSS
104-4024, GSS 064-4442, MDS uem0-043 and HDFS J223247.66-603335.9
in rest-frame B color (open circles). Dashed lines -- surface
brightness distribution of best fit model components using spatial
density distribution (\ref{eq2}). Solid line -- total surface
brightness.} \label{fig10}
\end{figure*}

It is possible to construct mass distribution models for galaxies
without assuming the existence of a DM component. The corresponding
models are given in Fig~\ref{fig10}. Now the calculated RCs start to
decrease significantly near or right after the last observed point
of rotation. The mass-to-light ratios for the disks are in this case
$M/L_B =$ 12., 1.8, 1.9 and 9.7, respectively. Thus, within the
presently available observations, the models without any dark halo
fit well in most cases with the observations and it is not possible
to determine the existence of a DM component in these four galaxies
on the basis of their kinematics. The mean mass-to-light ratio for
$z\sim 0.9$ galaxies (adding the two galaxies from Paper I) is now
$\langle M/L_B\rangle = 8.5$.

Figure~\ref{fig10} shows that RCs can be well fitted without DM.
However, we think that there are enough independent arguments in
support of the existence of DM and as a result, we think that
realistic models must include a DM component.

\begin{acknowledgements}
We thank the referee, Paolo Salucci, for a number of constructive
suggestions helping to improve the manuscript significantly. We
thank Dr. U. Haud for making available his programs for light
distribution model calculations. Surface photometry data used in
the present study are from STScI Archive and we thank the members
of observational proposals 5090 (PI E. Groth), 5369 (PI R.E.
Griffiths) and HDFS project team members for their work. We
acknowledge the financial support from the Estonian Science
Foundation (grant 4702).
\end{acknowledgements}

\end{document}